\documentclass[preprint,article,accept,moreauthors,pdftex]{mdpi} 

\firstpage{1} 
\pubvolume{1}
\issuenum{1}
\articlenumber{0}
\pubyear{2021}
\copyrightyear{2021}
\externaleditor{{Academic Editor: } 
} 
\datereceived{2 September 2021} 
\dateaccepted{15 September 2021} 
\datepublished{} 
\hreflink{https://doi.org/} 

\usepackage{changes}
\newcommand{\xadded}[1]{#1}
\newcommand{\xdeleted}[1]{}

\definecolor{burgundy}{rgb}{0.5, 0.0, 0.13}

\Title{A Bayesian Inference Framework for Gamma-Ray Burst Afterglow Properties}

\TitleCitation{A Bayesian Inference Framework for Gamma-Ray Burst Afterglow Properties}

\Author{En-Tzu Lin 
 $^{1,}$*\orcidA{}, Fergus Hayes $^{2}$\orcidB{}, Gavin P. Lamb $^{3}$\orcidC{}, Ik Siong Heng $^{2}$\orcidD{}, Albert K. H. Kong $^{1}$\orcidE{},\linebreak  Michael J. Williams $^{2}$\orcidF{}, Surojit Saha $^{1}$\orcidG{} and John Veitch $^{2}$\orcidH{}}

\AuthorNames{En-Tzu Lin, Fergus Hayes, Gavin P. Lamb, Ik Siong Heng, Albert K.H. Kong, Michael J. Williams, Surojit Saha, John Veitch}

\AuthorCitation{Lin, E.-T.; Hayes, F.;\linebreak Lamb, G.P.; Heng, I.S.; Kong, A.K.H.; Williams, M.J.; Saha, S.; Veitch, J.\linebreak}

\address{%
$^{1}$ \quad Institute of Astronomy, National Tsing Hua University, Hsinchu 300044, 
 Taiwan;\linebreak  akong@gapp.nthu.edu.tw (A.K.H.K.); suroastrophysics@gmail.com (S.S.)\\
$^{2}$ \quad SUPA, School of Physics and Astronomy, University of Glasgow, Glasgow G12 8QQ, UK; f.hayes.1@research.gla.ac.uk (F.H.); Ik.Heng@glasgow.ac.uk (I.S.H.); m.williams.4@research.gla.ac.uk (M.J.W.); John.Veitch@glasgow.ac.uk (J.V.)\\
$^{3}$ \quad School of Physics and Astronomy, University of Leicester, Leicester LE1 7RH, UK; gpl6@leicester.ac.uk \\
}

\corres{Correspondence: entzulin@gapp.nthu.edu.tw}

\abstract{In the field of multi-messenger astronomy, Bayesian inference is commonly adopted to compare the compatibility of models given the observed data. However, to describe a physical system like neutron star mergers and their associated gamma-ray burst (GRB) events, usually more than ten physical parameters are incorporated in the model. With such a complex model, likelihood evaluation for each Monte Carlo sampling point becomes a massive task and requires a significant amount of computational power. In this work, we perform quick parameter estimation on simulated GRB X-ray light curves using an interpolated physical GRB model. This is achieved by generating a grid of GRB afterglow light curves across the parameter space and replacing the likelihood with a simple interpolation function in the high-dimensional grid that stores all light curves. This framework, compared to the original method, leads to a $\sim$90$\times$ speedup per likelihood estimation. It will allow us to explore different jet models and enable fast model comparison in the future.}

\keyword{Bayesian inference; multi-messenger astronomy; GRB Afterglows} 

\begin{document}
\section{Introduction}
 
The gravitational wave emitted from a merging binary neutron star, GW170817~\citep{abbott2017gw170817}, followed swiftly by a Gamma-ray burst (GRB) of short duration known as GRB170817A~\citep{goldstein2017ordinary,savchenko2017integral} commenced the era of multi-messenger astrophysics (MMA). This is the first direct evidence showing neutron star mergers as the progenitor of short GRBs \citep{abbott2017_gw+grb}. The~union between electromagnetic (EM) and gravitational wave (GW) observations allow for astrophysical objects to be investigated with alternative probes, enabling the source properties to be studied under a new~light.


The coalescence of compact objects with at least one neutron star is of particular interest. By~the end of the third LIGO observing run (O3), one more binary neutron star (BNS) event,  GW190425~\citep{Abbott2020gw190425}, and two NS-BH mergers are reported~\citep{abbott2021observation}. However, GW170817 remains the only event with a confirmed GW-EM detection and has allowed a major leap in terms of its scientific~implications. 

A broadband search of the EM counterpart of GW170817 shows both thermal kilonova emission (AT2017gfo) generated from the radioactive decay of newly synthesized elements in the ejecta and a non-thermal synchrotron component from the relativistic jet~\citep{abbott2017multi}. The~latter is the GRB afterglow that typically shines in X-ray, optical, and radio wavelengths and can last for months to years following the GRB. For~the case of GRB170817A, the~first confirmed X-ray counterpart was observed by the Chandra X-ray Observatory and announced at around 9 days post-merger \citep{Troja2017}. Late-time monitoring of this event unveils the X-ray afterglow emission reaching its peak luminosity at around ~155 days~\cite{dobie2018,LIN2019} and continues to shine at more than 1000 days after the GRB \citep{Hajela2021,Troja2021,Troja2020}. 

Modeling the afterglow light curves would shed lights on the physical environment of the system, as~the afterglow peak time is dependent on the observers' viewing angle and the angular profile of the jet. Studies of the temporal behavior of GRB170817A afterglow suggest a structured jet seen off-axis, see, e.g., in~\cite{ghirlanda2019, lyman2018, lamb2019a, resmi2018, troja2018, Troja2020}. Understanding of the jet structure is essential to explain the observations of off-axis BNS events such as GRB170817A and uncovering the underlying physical mechanisms behind their production. With~better modelling of the jet structure, we are able to more confidently infer the rate of GRB production~\citep{tan2020jet}, and~the inference of the viewing angle of individual MMA events improves, allowing for tighter constraints on inference of the Hubble constant \citep{wang2021multimessenger,nakar2021afterglow}.

In this paper, we describe a semi-analytical GRB afterglow model that takes into account the effects of lateral spreading in Section~\ref{sec:model}. Based on this model, we develop a Bayesian framework that would allow quick parameter estimation and explain it in Section~\ref{sec:method}. We present the results with this methodology and discuss its future application in Section~\ref{sec:results}.

\section{GRB~Afterglow}\label{sec:model}

A broadband afterglow lasting hours-to-days typically follows a GRB, where this GRB afterglow is produced in the shock system that forms as the ultra-relativistic jet that emitted the GRB decelerates.
Using energy conservation for a spherical blastwave, the~dynamical behaviour of the decelerating system can be modeled, and~the change in Lorentz factor, d$\Gamma$, with~the change in swept-up mass found, see, e.g.,~in \cite{peer2012}.
The instantaneous $\Gamma$ and swept-up mass of the blastwave can be used to find the synchrotron emission responsible for the broadband GRB afterglow.
By following the method in~\cite{lamb2018}, a~single jetted outflow with a half opening angle, $\theta_j$, is split into multiple emission components and the dynamics and synchrotron flux, relative to the observers line-of-sight from each segment, can be found.
Summing across the multiple components at a given observer time, $t$, is equivalent to integrating across the equal arrival time surface for the jet, and~by allowing each component to have a unique energy and $\Gamma$, then the afterglow from complex jet structures can be found, see, e.g., in~\cite{lamb2017}.

This method naturally accounts for the edge-effect in a GRB afterglow, where the jet edge becomes visible for an on-axis observer when $\Gamma(t)<1/\theta_j$, and~the resulting jet-break in the lightcurve.
We further add the effects of lateral spreading in the jet by considering the maximum sound speed of each component and~the change in radius due to this lateral spreading (see in~\cite{lamb2018, lamb2021}).
With this method, the~afterglow from spreading and non-spreading jets, with~a defined structure profile can be found for observers at any viewing angle i.e.,~within the jet opening angle for cosmological GRBs, see, e.g.,~in \cite{lamb2019b}, and~at higher viewing angles for gravitational wave counterparts to neutron star mergers, see e.g.,~in\cite{lyman2018, lamb2019a}.

The semi-analytic approach used in generating the lightcurves divides the jet emission area over its radial and angular components into a grid of a given resolution.
The dynamics and flux are solved for each resolution element within the jet individually before summing to give the resultant lightcurve.
The duration of each afterglow iteration depends on the resolution, where for off-axis events the resolution can be set to low values >50, while for events with low viewing angle, wide opening angles or high $\Gamma$, the~required resolution becomes much higher, and~typically >10,000 to maintain the same level of accuracy.
Fitting algorithms can require 100s of thousands of individual model lightcurves for a reliable fit. In~our test model with four free parameters, it requires at least \mbox{$\geq$1 $\times10^{4}$ iterations} for a single analysis. A~model that takes several seconds, or~even minutes, per iteration is therefore not ideal.

\subsection*{Jet~Structuring}
For a power-law jet, the~energy and Lorentz factor distribution of the ejecta with respect to jet central axis is scaled as
$$
y(\theta)=\left\{
\begin{aligned}
  & 1 & & \text{if} & 0\le \theta \le \theta_{c},  \\
  & \Big( \frac{\theta}{\theta_{c}}\Big)^{-k} & &  \text{if} &\theta_{c} \le \theta \le \theta_{j},  \\
  & 0 & & \text{if} &\theta \ge \theta_{j}. 
\end{aligned}
\right.
$$

In this scenario, the~jet is parameterized by two characteristic angles, $\theta_{c}$ and $\theta_{j}$. Starting from the center of the jet, the~energy is uniformly distributed within the core, $\theta_{c}$. Outside of $\theta_{c}$, the~energy distribution follows a power-law decay with a sharp decline at the edge, $\theta_{j}$. This structure was proposed to account for the observed power-law decay in GRB afterglow light curves. Here, we assume the power-law index $k=2$ as used in \citet{lamb2017}.


\section{Parameter~Estimation}\label{sec:method}

The model described in Section~\ref{sec:model} is used to simulate X-ray afterglow light curves from 0.15 to 1000 days after the GRB event. Based on the X-ray fluxes, we generate random noise assuming a minimum signal-to-nose ratio (SNR) of 3. The~injection data are the sum of both the signal and the noise components. We perform the parameter estimation using Bayesian inference, from~which the posterior distribution of jet parameters is given by, according to Bayes theorem,
\begin{equation}
    p(\theta|d) = \frac{\mathcal{L}(d|\theta)\pi(\theta)}{\mathcal{Z}},
\end{equation}
where $\mathcal{L}(d|\theta)$ is the likelihood function of the data given a set of model parameters $\theta$, and $\pi(\theta)$ is the prior distribution for $\theta$ and $\mathcal{Z}$ is the evidence, or~the observed data. A~nested sampling algorithm is used to calculate the evidence and the posterior densities. We carry out the analysis with \textit{Bilby}, a~Python-based Bayesian inference library \citep{2019bilby}, and~the \textit{Dynesty} sampler 
 \citep{2020dynesty}. However, one of the issues with stochastic sampling over a complicated physical system is that the likelihood evaluation in the process are \mbox{computationally expensive. }

\begin{figure}[H]
\hspace{-0.2cm}\includegraphics[width=1\linewidth]{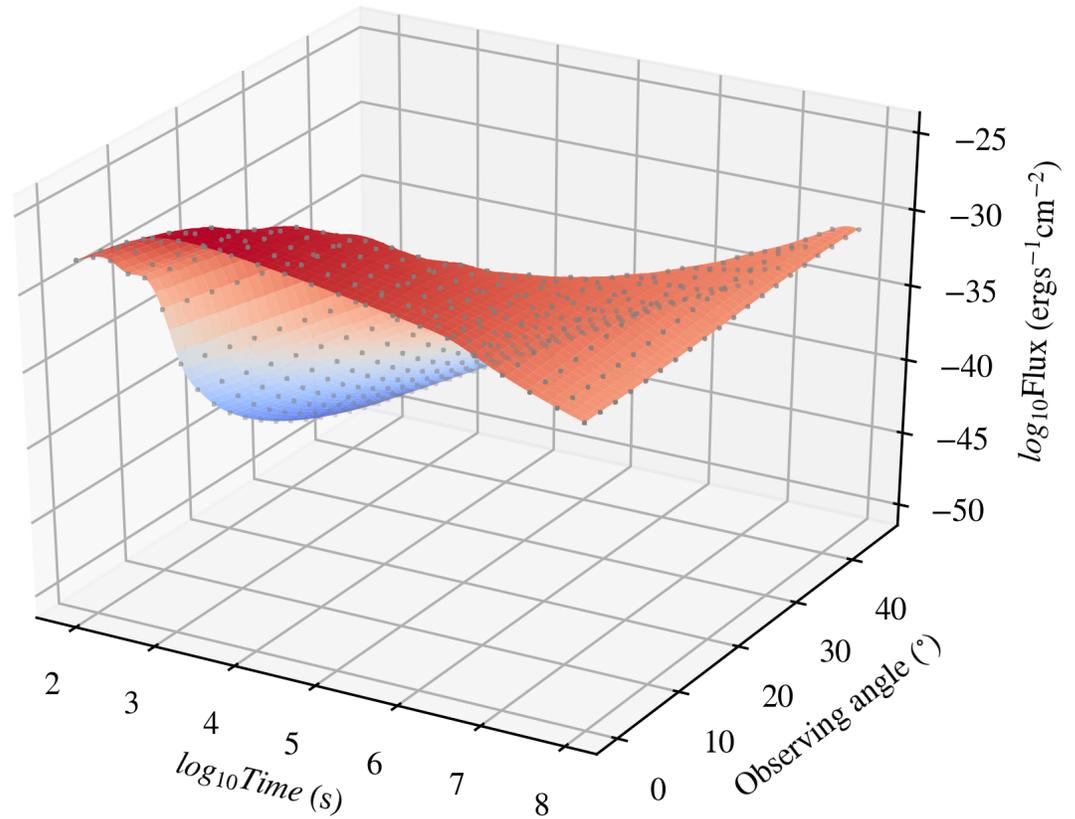}
\caption{Three-dimensional 
 illustration of afterglow light curves of a power-law jet seen from $\theta_{obs} = 0^{\circ}$ to $\theta_{obs} = 45^{\circ}$. Gray dots are data stored in our 4D grid. The~open-angle of the jet is \mbox{$\theta_{j}=0.2$ (rad)} and the core width $\theta_{c} = 0.15$ (rad). Other fixed parameters include \mbox{$\epsilon_{0} = 2\times10^{52}$ ergs$^{-1}$,} \mbox{$n=0.001$ cm$^{-3}$,} $p=2.15$, $\epsilon_{B}=0.01$, $\epsilon_{e}=0.1$ and $d=40$ Mpc.}
\label{fig:box}
\end{figure}

In order to increase the efficiency in investigating different jet models, we employ an interpolated GRB model and substitute the original likelihood function with it. This is done by first simulating each model parameter across a certain range in the parameter space and~storing the resulting light curves in a multidimensional grid. The~detailed configuration of the grid is specified below:

\begin{itemize}
    \item $log_{10}\epsilon_{0}\sim(50, 53)$
    \item $\theta_{j}\sim(0,  30^{\circ}$)
    \item $\theta_{c}\sim(0,  30^{\circ}$)
    \item $\theta_{obs}\sim(0,  45^{\circ}$), 
\end{itemize}
\textls[-25]{where 
$\theta_{j}$ is the jet half-opening angle, $\theta_{c}$ is the core width of a structured jet, $\theta_{obs}$ is the angle between the observer and the jet central axis, and~$\epsilon_{0}$ is the isotropic equivalent energy of the jet. For~each dimension, the~parameter range is linearly cut into subsets. We increase the density of splits for parameters that \xadded{have higher impacts to the resulting light curves} \xdeleted{cause the most variation to the light curves}. The~total number of entries in our 4D Cartesian grid is then $\epsilon_{0}\times \theta_{j}\times \theta_{c}\times\theta_{obs} = 40\times 50 \times40 \times50 =$ 4,000,000.} \textls[-15]{Figure~\ref{fig:box} \xadded{shows} \xdeleted{is a demonstration of} how afterglow light curves distribute in the three-dimensional space for an observer at different viewing angles. When the inclination angle increases, the~afterglow flux peaks at later times. Jet parameters in this dataset are $\theta_{j}=0.2$ rad ($\sim$$11.5^{\circ}$), \mbox{$\theta_{c} = 0.15$ rad}} ($\sim$$8.6^{\circ}$) and $\epsilon_{0} = 2\times10^{52}$ ergs$^{-1}$.

To put the emphasis on the jet structure and save computational power, we keep the rest of the afterglow parameters constant throughout the simulation. The~fixed parameters include the initial bulk Lorentz factor $\Gamma = 100$, the~ambient particle number density \mbox{$n=0.001$ cm$^{-3}$,} the shock accelerated election index $p=2.15$, the~microphysical parameters $\epsilon_{B}=0.01$ and $\epsilon_{e}=0.1$ \xadded{\citep{lamb2017}}, and~the luminosity distance $d=40$ Mpc \xadded{\citep{abbott2017gw170817}} for a GW170817-like source   \xdeleted{\citep{Lazzati2017a}}. The~resolution in the jet model is set to 50 for all light curves. 

\begin{figure}[H]
\includegraphics[width=1\linewidth]{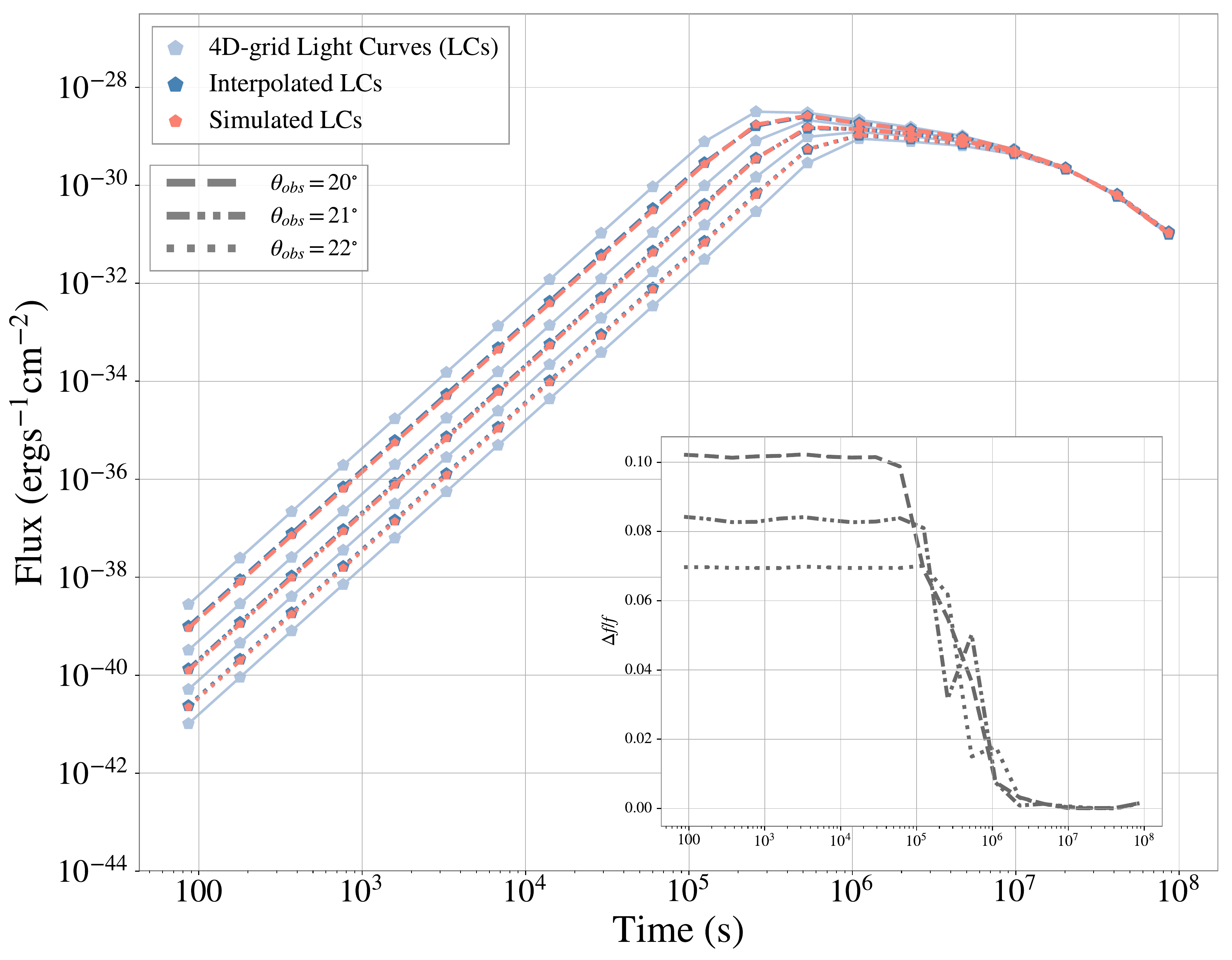}
\caption{A comparison between the interpolated \xadded{(dark blue)} and simulated light curves \xadded{(red) at observation angles of $\theta_{obs} = 20^{\circ}, 21^{\circ}$ and $22^{\circ}$. These observing angles lie in the midpoints of the 4D-grid of simulated light curves (light blue) used to perform the interpolation. The~inset shows the fractional difference between the interpolated light curves and the simulated light curves.} \xdeleted{Light blue dots are light curves stored in the 4D-grid where as dark blues are interpolated light curves at $\theta_{obs} = 20^{\circ}, 21^{\circ}$ and $22^{\circ}$. Given the same set of values, we use the original GRB model to generate the simulated light curves (red). The~inset shows the percentage by which the interpolated light curves deviate from the theoretical values.} Note that all parameter values are same as Figure~\ref{fig:box} except for a slightly wider jet with $\theta_{j} = 0.3$ rad ($\sim$$17^{\circ}$).}

\label{fig:lc}
\end{figure}

We adopt uniform priors with upper and lower bounds the same as the simulation range for every parameter of interest. At~each Monte Carlo sampling point, instead of calculating the likelihood from the original model, it performs a linear interpolation between the adjacent parameter values in the grid. A~comparison between the interpolated and theoretical light curves with same parameter values is shown in Figure~\ref{fig:lc}. The~largest difference appears at observing angles near the jet edge, i.e.,~slightly off-axis case with \textasciitilde10\% deviation. As~the geometry between the jet and observers affects the light curves the most, we set the numbers of splits for $\theta_{j}$ and $\theta_{obs}$ larger than other parameters. Increasing the density of splits in the grid would raise the accuracy but the simulation time also dramatically grows. For~post-peak observing epochs and the rest of off-axis scenarios, the~interpolated light curves well resemble the theoretical values with a <10\% deviation. As our input data has a SNR value of 3, the~error caused by the lack of entries in the interpolation process is not the dominant source of uncertainty.

\unskip

\section{Results and~Discussion}\label{sec:results}

\begin{figure}[H]
\hspace{-0.2cm}\includegraphics[width=0.7\linewidth]{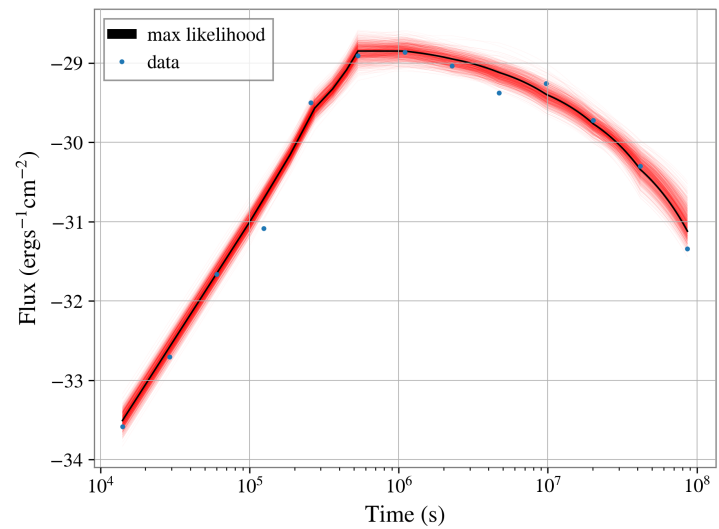}
\caption{Injection data points (blue) overlaid with the best-fit power-law jet model (black solid line). Red lines are 1000 random draws from the posterior samples. 
\label{fig:posterior}}
\end{figure}

\begin{figure}[H]
\includegraphics[width=0.92\linewidth]{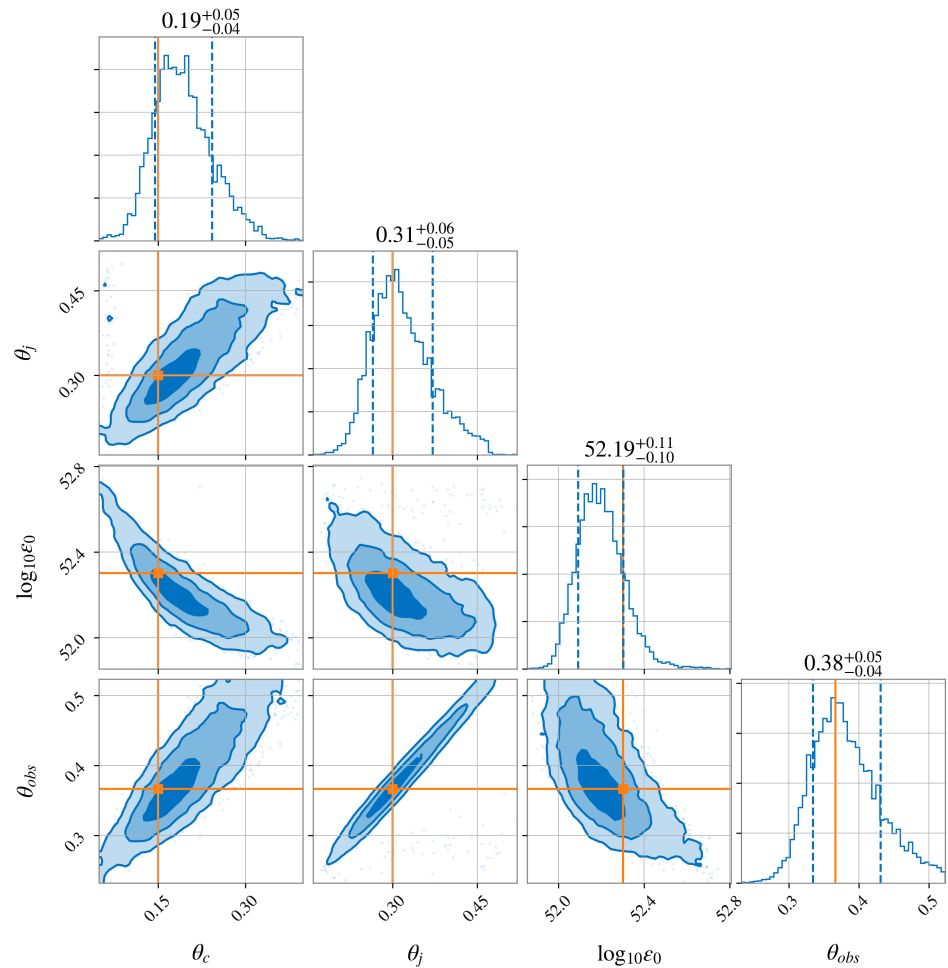}
\caption{Probability density distribution of parameters in a power-law structured jet model.  Orange vertical lines indicate the injected values from which the input data was simulated. 
\label{fig:corner}}
\end{figure}

The input data, produced with the method described in Section~\ref{sec:method}, are the X-ray light curve generated from a GRB jet with a power-law profile. The~width of the jet is set to $\theta_{j} = 0.3$ rad with the core angle being $\theta_{c}=0.15$ rad. Considering an off-axis observer, we set the viewing angle $\theta_{obs} = 0.36$ rad ($21^{\circ}$) and the isotropic energy is \mbox{$\epsilon_{0}= 2\times10^{52}$erg s$^{-1}$.} The \xadded{remaining} \xdeleted{rest of the} parameters have the same values as those used to simulate the 4D grid. This light curve is composed of the 0.4 keV fluxes at observing times from 0.16 days to 1000 days post-event and visualized as blue dots in Figure~\ref{fig:posterior}. After~incorporating the interpolated model in the sampling process, the~time required for calculating the likelihood per iteration reduces from 7.78 seconds to $\sim$$0.09$ seconds with only 1 CPU. 
The overall run time with this method decreases by $\sim$$90$ times. 
Figure~\ref{fig:corner} shows \xdeleted{the} the joint posterior of the four parameters mentioned above. The~injection values are well recovered within the 68\% confidence interval. This serves as a proof-of-principle that this method is not only efficient, but~also capable of capturing the jet features if the afterglow emission is intrinsically produced by such a~system. 

In this work, we perform an analysis on simulated light curves with a uniform signal-to-noise ratio throughout all observing epochs. In~reality, for~the same patch of sky and the same observational instrument, the~background fluence would remain at a similar scale. The~temporal evolution of a GRB afterglow, which arises until reaching the peak luminosity at $t_{peak}$ and then declines with time, yields a varying observational significance. The~X-ray emission from the position of GW170817, except~for the first observation at 2.3 days, reports a significance of $\geq$$3\sigma$ at all epochs. We set all SNR equaling to the detection limit for a conservative estimate. Recent studies show \xadded{that} the X-ray counterpart of GW170817 is still bright after 1200 days \citep{Hajela2021, Troja2021}. The~$3\sigma$ excess compared to the prediction from an off-axis structured jet poses a challenge to previous agreements from multi-wavelength studies. It also makes modeling the afterglow of GRB170817 worthwhile even after four years as new evidence is still emerging.

\vspace{-12pt}
  
On the other hand, we use relatively wide\xdeleted{,} uniform priors for jet structure parameters in the analysis. One of the strengths of multi-messenger astronomy is that we can combine the information obtained from various approaches and place tighter constraints on the source properties. As~gravitational wave observation provides an independent measure of viewing angle and luminosity distance, the~posteriors from GW \xadded{parameter estimation} \xdeleted{PE} can be passed to the afterglow analysis as prior function and thus increase the accuracy in search of the GRB afterglow~parameters.

In this paper, we present the validity of a Bayesian framework combined with an interpolated GRB model. It effectively lessens the computational cost while being capable of constraining jet structure parameters in low SNR circumstances. In~the upcoming era of a global gravitational wave detector network (LIGO-VIRGO-KAGRA) and new EM observatories like the Gravitational Wave High-energy Electromagnetic Counterpart All-sky Monitor (GECAM), it is foreseeable that we will find more GW-EM counterparts in the near future. A~fast, functional parameter estimation will be beneficial for modeling the GRB afterglow light curves. With~this methodology, more jet models, e.g.,~Gaussian or Top-hat jets, will be investigated and its application on the real data of GW170817, the~only event so far with GW-EM detection, will be included in our future~works.


\vspace{6pt}
\authorcontributions{Conceptualization, E.-T.L., F.H. and I.S.H.; Formal analysis/ Investigation/Visualization, E.-T.L.; Methodology, E.-T.L., F.H. and G.P.L.; Writing – original draft, E.-T.L, F.H. and G.P.L.; Writing – review and editing, A.K.H.K. and S.S.. All authors have read and agreed to the published version of the manuscript. 
}

\funding{This work is supported by the Ministry of Science and Technology of Taiwan through grant~109-2628-M-007-005-RSP.}

\conflictsofinterest{The authors declare no conflict of interest. 
} 



\end{paracol}
\reftitle{References}


\externalbibliography{yes}
\bibliography{references.bib}


\end{document}